# Auxiliary Chemical Geothermometers Applied to Waters from some East African Rift Areas (Djibouti, Ethiopia, Kenya) for Geothermal Exploration

**Bernard Sanjuan**

**BRGM, Georesource Division, 3 avenue C. Guillemin, 45060 Orléans, France**
*b.sanjuan@brgm.fr*

## ABSTRACT

If the temperature values of geothermal reservoirs given by classical chemical geothermometers applied on thermal waters are relatively divergent, it is often difficult to estimate the temperatures of these reservoirs with sufficient accuracy for geothermal exploration, before drilling operations. In this case, some auxiliary chemical geothermometers such as Na-Li, Mg-Li, Na-Rb, Na-Cs, K-Sr, K-F, K-Mn, K-Fe, K-W, etc., existing in the literature can be useful tools to help estimating these temperatures. However, previous studies have shown these geothermometers are not only dependent on temperature, but also on other parameters such as the nature of the reservoir rocks and minerals, the fluid salinity, etc. Consequently, they must be used with caution. Another problem for the use of these geothermometers is that Li, Sr, Cs, Rb, Mn, Fe and W are under the form of trace elements in numerous geothermal waters and that it is often difficult to find analyses of these elements in the literature.

In this study, we test some of these auxiliary geothermometers on several waters selected from literature data relative to East African Rift (EAR) geothermal areas, in Republic of Djibouti, Ethiopia and Kenya, where temperature values of deep geothermal reservoirs have been measured into wells or estimated with certainty using chemical classical geothermometers applied on thermal waters. The comparative results are discussed and allow bringing conclusions and recommendations. Among the tested auxiliary geothermometers, this work shows that the Na-Li thermometric relationship defined for the dilute waters from high-temperature volcanic geothermal areas of Iceland is one of the most relevant to estimate the temperatures of deep reservoirs for waters from several geothermal areas from the Republic of Djibouti, Ethiopia and Kenya. In previous studies, this Na-Li thermometric relationship had also given good estimations of reservoir temperatures for high-temperature (T $\geq$ 300°C) borehole dilute waters from the Los Humeros geothermal area, in Mexico, in volcanic environment. The use of other auxiliary geothermometers such as K-F, K-Sr, Na-Rb and Na-Cs (when F, Sr, Rb and Cs have been analysed) can be also relevant in some cases.

## 1. Introduction

Several studies about auxiliary chemical geothermometers such as Na-Li, Mg-Li, Na-Rb, Na-Cs, K-Sr, K-F, K-Mn, K-Fe, K-W have been carried out in the literature for geothermal prospecting in different geological environments, Fouillac and Michard(1981); Kharakha and Mariner(1982); Kharakha et al.(1989); Michard(1990); Sanjuan et al.(2014; 2016a, b). These geothermometers can be very useful when the classical chemical geothermometers such as silica, Na-K, Na-K-Ca, and K-Mg do not give a concordant estimation of reservoir temperature. However, as they do not only depend on temperature, but also on other parameters such as the nature of the reservoir rocks and minerals, their degree of alteration, the fluid chemistry and salinity, etc. (Sanjuan et al., 2014; 2022), they must be used with caution. It is essential to well define the environment in which these geothermometers will be applied before their use.

If the different Na-Li thermometer relationships for thermal waters in contact with granite and volcanic rocks, Fouillac and Michard(1981); Michard(1990); Sanjuan et al.(2014) or for geothermal brines in oil- and geothermal field sedimentary basins, Kharakha and Mariner(1982); Kharakha et al.(1989) are the most known and used in the literature, very few studies have been carried out using the other auxiliary chemical geothermometers. Michard (1990) showed that chemical geothermometers such as Na-Cs, Na-Rb, K-Sr, K-Mn, K-Fe, K-F, K-W... could also be used for geothermal exploration concerning dilute waters discharged from granite reservoirs between 25 and 150°C in more than sixty areas from Europe (France, Italy, Spain, Bulgaria, Sweden). Sanjuan et al. (2016a, b) defined three new Na-Rb, Na-Cs and K-Sr thermometric relationships using 20 hot natural brines from granite and sedimentary reservoirs, mainly located in the Upper Rhine Graben, France and Germany (70-230°C), apart two which were at Salton Sea, in the imperial Valley, USA (320-340°C).

The main objective of this paper is to test some of these auxiliary geothermometers on several waters selected from literature data relative to the East African Rift (EAR) geothermal areas, in Republic of Djibouti, Ethiopia and Kenya, where temperature values of deep geothermal reservoirs have been measured into wells or estimated using chemical classical geothermometers applied on thermal waters. The comparative results will be discussed and allow bringing conclusions and recommendations about the use of these new tools for geothermal exploration in EAR geothermal areas.

## 2. EAR geothermal areas selected in this study

The East African Rift System (EARS) is one of the most important volcano-tectonically active regions where heat energy from the Earth's interior escapes to the surface in the form of volcanic eruptions and the upwelling of heat by hot springs and fumaroles. Therefore, the EARS appears to possess a remarkable geothermal potential, Pürschel et al.( 2013). However, at the present, only Kenya has been able to develop a considerable industrial production of geothermal power (1193-installed MWe in 2020; Huttrer, 2020). There is no geothermal power station in the Republic of Djibouti, despite numerous works of geothermal exploration. Several geothermal areas were selected from studies carried out in volcanic environment in Republic of Djibouti, Ethiopia and Kenya. As the Republic of Djibouti was the country where more geochemical data required for this study were found, the latter were more detailed.

## 2.1 Republic of Djibouti

The Republic of Djibouti is located in the EARS, where three major extensional structures (Red Sea, EAR and Gulf of Aden Rift systems) join to form the Afar depression (Fig. 1A), which is characterized by thinned continental crust.

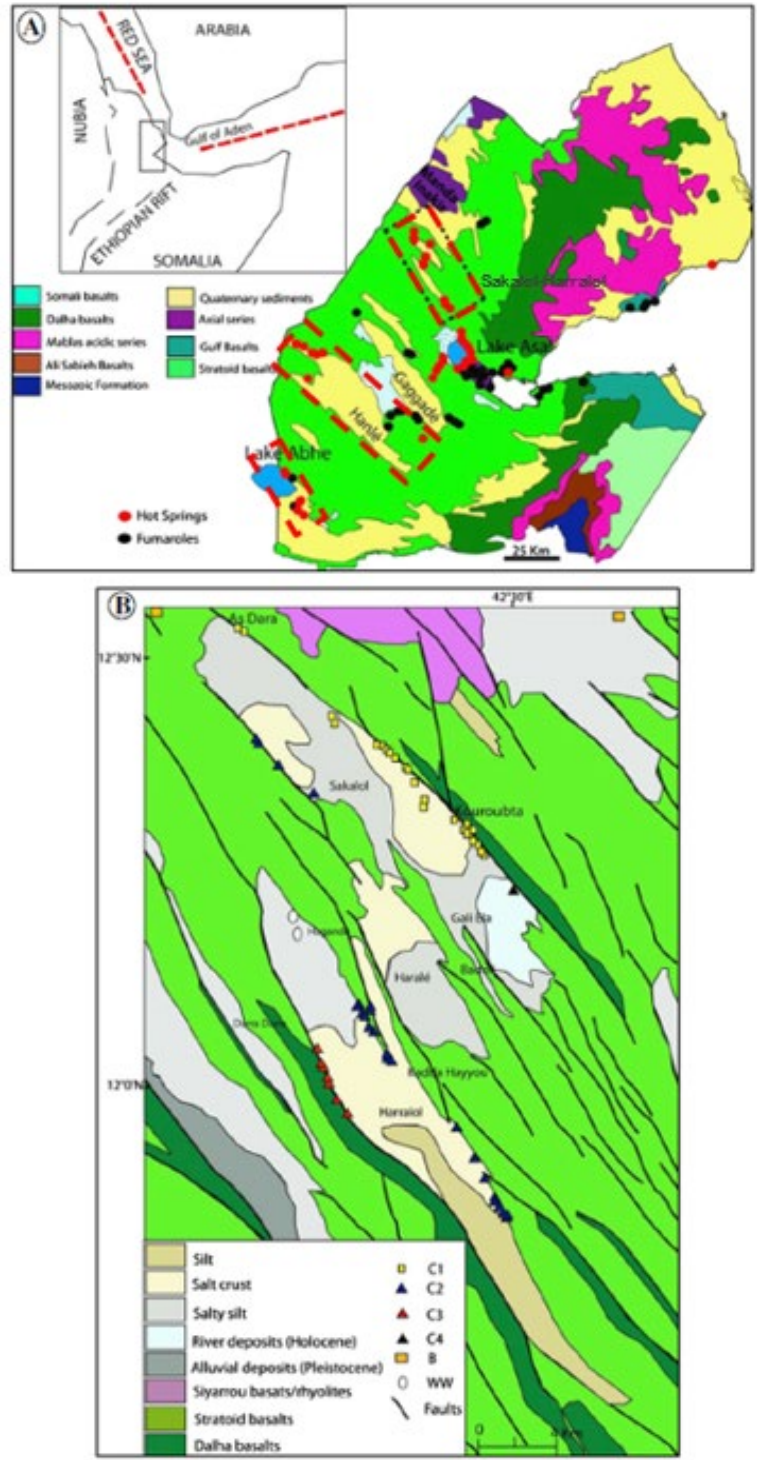

**Figure 1: A): A schematic geological map of the Republic of Djibouti (SE Afar Rift) with hydrothermal activity of the Republic of Djibouti. (B): A schematic geological map of the Sakalol-Harralol geothermal area with thermal springs organized in 4 clusters (C1, C2, C3 and C4). In the inset: schematic map of the Afar Depression with the location of Djibouti (black rectangle) (extracted from Awaleh *et al.*(2017).**

This depression is bounded by large escarpments to the west and to the south, and by the Danakil Alps to the northeast, Zan *et al.*(1990). Almost all the Republic of Djibouti is covered by volcanic rocks, mainly represented by basalts, and thermal manifestations are widespread (Fig. 1A).

The areas selected for this work were the Sakalol-Harralol Geothermal Field, mainly studied by Awaleh *et al.* (2017), the Hanlé-Gaggadé Geothermal Field, personal data (1988); Awaleh *et al.*(2020) and the Lake Abhe Geothermal Field, Awaleh *et al.*(2015). These areas have been less studied than the high-temperature Asal-North Ghoubbet area where several geothermal wells have been drilled. For this last area and the Tadjourah and Obock areas where the geothermal fluids are mainly derived from seawater-basalt interaction processes at high-temperatures ($\geq$ 160°C) as well as for the Reykjanes area in Iceland, Sanjuan *et al.* (2014) already determined a specific Na-Li thermometric relationship.

### a) Sakalol-Harralol Geothermal Field (SHGF)

The SHGF (about 200 km$^2$) is one of the largest geothermal fields of the Republic of Djibouti, which is located in the north of the EARS and within the northwestern portion of the Ghoubbet-Asal, which forms an accretionary rift segment penetrating the Afar depression (Fig. 1A). This rift is one of two emerged oceanic ridges in the world, with the other being Iceland. As noted by Awaleh *et al.* (2017), similar geological features are the Manda Inakir, the Manda Hararo and the Erta Ale.

Detailed geochemical investigations were carried out for the first time by Awaleh *et al.* (2017) on cold groundwaters (well and borehole waters) and almost all thermal waters (86 thermal springs) from this area (Fig. 1B). The temperature of these thermal springs at surface are ranging from 38 to 78°C and their pH values moderately alkaline from 7.15 to 8.95.

Using common statistical analyses and the chemical composition of the waters described in terms of relative concentrations of the main major species, Awaleh *et al.* (2017) distinguished four clusters of thermal waters in the SHGF (Fig. 1B). The TDS values are increasing from the clusters C1 (minimum TDS of 715 mg/l) to C4 (maximum TDS of 13265 mg/l). Except the cluster C1 for which the waters are mostly of the Na-Cl-HCO$_3$-SO$_4$ type, the geothermal waters of the other clusters are Na-Cl type. The cluster C4 is constituted of the most saline and hottest water at surface (T = 77.7°C).Using different geothermometric approaches, Awaleh *et al.* (2017) gave a temperature range estimation for the deep geothermal reservoir in the Sakalol-Harralol area of about 120-160°C, with a mean deep temperature of 143°C. In addition, according to previous hydrological studies, they noted that the presence of a shallow aquifer at 110°C could be also likely.

### b) Hanlé-Gaggadé Geothermal Field (HGGF)

Within the EARS, the HGGF is located southwestwards of the Asal Rift (Fig. 1A). The Hanlé and Gaggadé plains are two of the many tectonic depressions lying parallel to the Asal Rift, between this structure and Lake Abhe, Zan *et al.*(1990). These two half-grabens and the corresponding intra-basins are 18 km and 10 km-wide, respectively. They have a similar orientation, which is northwest-southeast, and are bounded by a prominent system of master and secondary normal faults, Awaleh *et al.*(2020). The hanging wall of the Hanlé half-graben is locally disrupted by the emplacement of a large acid intrusion, which forms the domed Baba Alou relief culminating at approximately 972 m and is further dissected by high-angle faults. The Upper Stratoid series (2.2-1.8 Ma) is the main cartographic unit and the basaltic component covers most of its surface (Fig. 1A).

The data selected for this study are relative to 10 thermal springs and the deep well H1 in the Hanlé plain, and 2 thermal springs in the Gaggadé plain, personal data (1988); Table 1). The geochemical data from the study of Awaleh et al. (2020) were also selected. The temperature of these thermal springs at surface are ranging from 38 to 70°C. Two geothermal wells (H1 and H2) were drilled in the Hanlé plain in 1987, with depths of 1623 m and 2038 m, respectively, Zan et al.(1990). Maximum temperatures of 72°C and 124°C were recorded in each of these wells. As noted by Zan et al. (1990), temperature logs run in wells H1 and H2 show the Hanlé plain is a system in which temperature seems to be controlled by groundwater circulation (down to 800 m in H1 and to about 1000 m in H2). It was found that the local temperature maximum at shallow depth is connected to aquifers flowing in a lateral direction. The zone with an almost constant temperature, from about 400 to 1000 m in well H2, could be related to the local thermal anomaly generated by the ascent of hot fluids to the Garrabbays fumaroles. All these thermal waters are moderately alkaline with pH values ranging from 7.62 to 8.86. Their TDS values are ranging from 748 mg/l to 2910 mg/l. Except for the Nɛinlé thermal water G1, which is Na-HCO$_3$-Cl type, all the other waters are Na-Cl type.

**Table 1: Chemical composition of geothermal waters from the HGGF, personal data (1988).**

| Sample | Sample name | Latitude | Longitude | T °C | pH | TDS g/l | Na mg/l | K mg/l | Ca mg/l | Mg mg/l | Cl mg/l | HCO$_3$ mg/l | SO$_4$ mg/l | SiO$_2$ mg/l | Br mg/l | B mg/l | Li mg/l | Sr mg/l | Rb mg/l | Cs mg/l |
|---|---|---|---|---|---|---|---|---|---|---|---|---|---|---|---|---|---|---|---|---|
| H1 | Galafi borehole | 11°42'12.30" | 41°51'0.50" | 50.0 | 8.15 | 1.9 | 621 | 37.5 | 3.81 | 0.75 | 635 | 200 | 280 | 120 | 3.36 | 0.627 | 0.031 | 0.085 | 0.071 | 0.082 |
| H2 | Boukboukto | 11°41'8.25" | 41°52'0.84" | 48.0 | 8.01 | 1.9 | 621 | 37.5 | 4.29 | 0.80 | 638 | 211 | 275 | 103 | 3.40 | 0.627 | 0.026 | 0.093 | 0.048 | 0.077 |
| H3 | ɛasa Mayeb | | | 43.0 | 8.27 | 1.9 | 635 | 37.5 | 3.69 | 1.07 | 624 | 220 | 275 | 84 | 3.32 | 0.627 | 0.021 | 0.101 | 0.036 | 0.072 |
| H4 | Dåli | 11°39'0.00" | 41°55'4.74" | 38.5 | 8.01 | 2.9 | 1039 | 49.7 | 6.77 | 2.67 | 1064 | 223 | 412 | 63 | 5.75 | 1.103 | 0.012 | 0.145 | 0.050 | 0.070 |
| H5 | ɛaddara | 11°38'36.56" | 41°55'30.73" | 39.0 | 8.20 | 2.9 | 1048 | 45.4 | 5.41 | 1.97 | 1064 | 272 | 407 | 66 | 5.75 | 1.146 | 0.009 | 0.136 | 0.049 | 0.080 |
| H6 | Dahotto | 11°37'31.81" | 41°57'3.73" | 43.0 | 7.83 | 2.6 | 910 | 25.8 | 8.50 | 4.47 | 964 | 229 | 342 | 70 | 5.23 | 0.908 | 0.011 | 0.129 | 0.000 | 0.064 |
| H7 | Minkille | 11°39'16.15" | 41°56'59.19" | 58.0 | 7.62 | 2.3 | 726 | 14.5 | 25.17 | 3.55 | 922 | 73 | 456 | 87 | 4.67 | 0.951 | 0.036 | 0.423 | 0.041 | 0.070 |
| H8 | Daggirou | 11°38'30.28" | 41°58'37.13" | 41.0 | 7.82 | 2.4 | 809 | 30.1 | 12.67 | 4.03 | 957 | 77 | 329 | 72 | 4.95 | 0.822 | 0.015 | 0.117 | 0.000 | 0.068 |
| H9 | Oudgini | 11°30'47.17" | 41°56'14.88" | 40.5 | 7.89 | 1.9 | 644 | 21.5 | 3.05 | 4.23 | 798 | 147 | 226 | 79 | 4.27 | 0.724 | 0.016 | 0.158 | 0.043 | 0.064 |
| H10 | ɛagna | 11°33'51.05" | 41°54'32.18" | 40.0 | 7.81 | 1.9 | 602 | 26.6 | 6.49 | 2.84 | 709 | 193 | 244 | 74 | 3.80 | 0.714 | 0.012 | 0.077 | | |
| H11 | Hanlé 1 well | | | 72.0 | 8.86 | 1.5 | 483 | 18.6 | 8.02 | | 491 | 206 | 202 | 56 | 0.08 | 0.000 | 0.028 | | | |
| G1 | Nɛinle | | | 42.0 | 7.88 | 748 | 136 | 5.5 | 15.43 | 2.28 | 138 | 272 | 122 | 58 | 0.32 | 0.238 | 0.017 | 0.578 | 0.007 | 0.011 |
| G2 | Nɛinle | | | 70.0 | 7.85 | 1203 | 340 | 18.4 | 15.83 | 2.14 | 383 | 114 | 201 | 129 | 1.72 | 0.422 | 0.031 | 0.425 | 0.046 | 0.071 |

For the G1 and G2 Nɛinle thermal waters, we estimated the reservoir temperature at about 140°C, using the Na-K and silica-chalcedony geothermometers, Arnorsson et al. (1983). Awaleh et al. (2020) proposed a conceptual model of the Hanlé-Gaggadé system, with a mean temperature of 145 ± 15°C for the main geothermal reservoir, which would be located at a maximum depth of 2400 m. This temperature was estimated using different geothermometric approaches.

### c) Lake Abhe Geothermal Field (LAGF)

This geothermal field is located in the Southwestern region of the Republic of Djibouti, on the border with Ethiopia (Fig. 1A). It occurs within a rift basin filled with Pliocene-Quaternary volcanic rocks (mainly basalt) and lacustrine sediments. The lake sediment floor is underlain by a thick sequence of Stratoid basalts dated between 4 and 1 Ma. During the Plio-pleistocene, these basalts were dislocated by extensional faulting, with fault-scarps exceeding 1000 m in height in some areas, allowing the development of deep lakes in the Central Afar and favoring groundwater movement between the different basins like Abhe, Dobi-Hanlé and Asal.

The Lake Abhe area is particularly rich with surface hydrothermal features, including fumaroles, hot and warm springs and hydrothermal chimney structures, some of which discharge hot steam at their apex. These are aligned WNW-ESE, parallel to the regional extensional fault network. The surface hydrothermal manifestations are spread over an area of about 100 km$^2$.

The sixteen hot waters studied by Awaleh *et al.* (2015) were selected for this study (SHC1-SHC7 group located at the north of the GHC1-GHC9 group). These Na-Cl thermal waters have temperatures at surface ranging from 71 to 99.7°C and pH values varying from 7.61 to 8.79. The hot waters from the SHC1-SHC7 group have TDS values higher than those from the GHC1-GHC9 group (3466 to 3795 mg/l against 1918 to 2236 mg/l). The different geothermometric approaches used by Awaleh *et al.* (2015) estimated a temperature range of the deep geothermal reservoir of 120-160°C. In spite of the relatively wide range, the three different approaches led to a same mean value of about 135°C.

### 2.2 Ethiopia

Ethiopia is located on the geothermally active East African Rift System (Afar depression and Main Ethiopian Rift, MER; Fig. 2) and therefore has an abundance of sites that are prospective for generation of power. Twenty-four such locations are claimed as is a potential for ultimate generation of 10,000 MWe, Huttrer(2020). The current installed capacity is 7.3 MWe derived from the Aluto Langano field. It is located in the southern part of the rift (Fig. 2). The prospects most advanced include Tendaho and its associated Alalobeda area, Shalla Abiata, Butajira, Meteka, Corbetti, and Tulu Moye (Fig. 2). Exploration at several of these has recorded temperatures greater than 200°C.

The Eastern branch, which forms the Ethiopian and Kenyan rifts, marks the boundary between the Nubian and Somalian plates. Through its high regional heat flow due to an underlying basic upper mantle intrusion beneath the thinned crust, it exhibits, by far, the most extensive geothermal resources. Most of the widespread geothermal activity, manifesting itself in the form of numerous hot springs, fumaroles and hydrothermal alteration, is located in the MER and in the Afar depression (Fig. 2). Recent volcanic activity is characterized by Mid-Ocean Ridge Basalt (MORB)-like fissural eruptions. Faulting is a typical and dyke induced.

Through the narrow fissures, which have penetrated the crust in the rift axis, the basaltic magma erupted to the surface and formed chains of cones, dykes and sills. In the MER, the water circulates along the dominating NS to NNE trending fault system within the axial valley and originates from the rift flanks. The heat source is provided by dykes and central magma chambers. The major aquifers in this zone are fractured, interlayered basalts and ignimbrites.

For this study, the geothermal waters were selected from four main works carried out by Endeshaw (1988), AQUATER (1996), Pürschel *et al.* (2013) and Minissale *et al.* (2017). Pürschel *et al.* (2013) mainly worked on hot springs from three geothermal areas (Dofan-Fantale, Gergede-Sodere and Aluto-Langano; (Fig. 2). The Gergede-Sodere thermal springs emerge along the young faults on axial part of the rift, whereas the Dofan-Fantale and Aluto-Langano hot springs are associated with active volcanic centers.

Among the waters selected in this study, the geothermal waters Bulga 1 from the Dofan-Fantale area, and Sodere 1 and 2 from the Gergede-Sodere area, are Na-HCO$_3$ type. The Langano 1, Langano 2, L. Spring 84, and L. Spring 10 waters from the Aluto-Langano geothermal field, are also Na-HCO$_3$ type. The water from the geothermal well LA-4 (1987) is Na-HCO$_3$ type and has a TDS value close to 4 g/l.

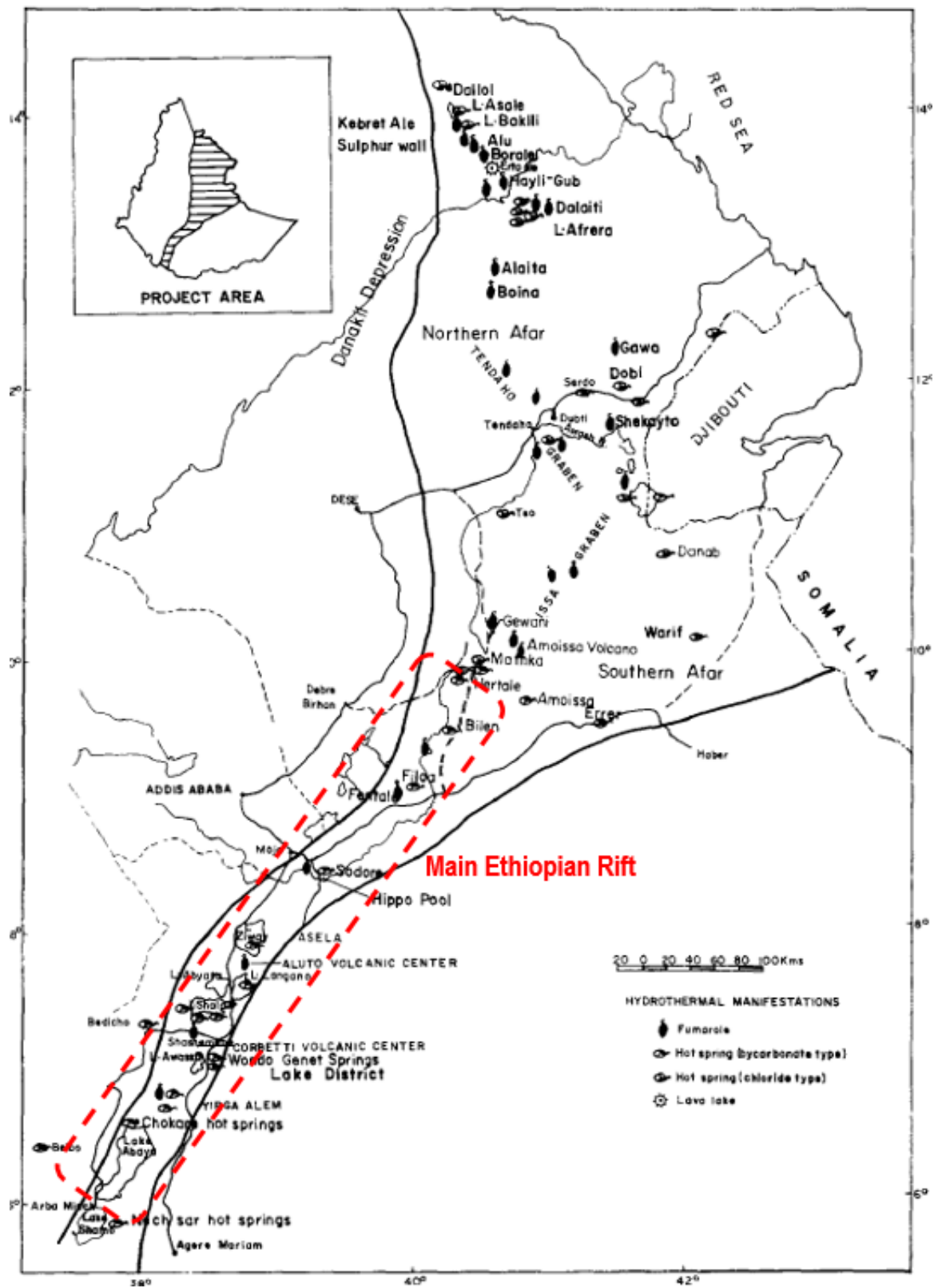

**Figure 2: Location map of the Ethiopian Rift Valley showing the main geothermal fields and hydrothermal manifestations (extracted from Endeshaw, 1988).**

Using different chemical geothermometers, Pürschel *et al.* (2013) concluded that the Na-K and Na-K-Ca geothermometers provided the most reliable subsurface temperature estimates with 185±20°C for the investigated hot spring samples and 260±15°C for the LA-4 fluid sample.

From the study carried out by AQUATER (1996) in the Tendaho geothermal area, the geothermal waters from the deep wells TD-1, TD-2 and TD-4 were selected for this study. After their drilling, these wells yielded a temperature of over 250°C.

Ten geothermal waters from different areas were selected in the two other studies: Dallol, Lake Afrera, Hertale, Bilen, Filweha and Lake Abaya 6 waters (Fig. 2; Endeshaw, 1988) and Bilate, Lake Abaya 6, 8, and Dimtu well waters from the northern Lake Abaya area, Minissale *et al.*(2017). In their study, Minissale *et al.* (2017) concluded that the application of geothermometric techniques in the liquid and the gas phases suggests the presence of a deep reservoir in which the fluids equilibrated at a maximum temperature of approximately 180°C in the northern Lake Abaya area (for the Bilate sample, for example). For the Lake Abaya fluid samples, the temperature reservoir is estimated at 260°C and at 150°C for the Dimtu well water, using classical geothermometers. For the other study, Endeshaw(1988), the use of different geothermometers give estimations of reservoir temperatures of about 100°C for the Filweha water, 110°C for the Dallol water, 140°C for the Bilen water, 160°C for the Hertale water and 180°C for the Lake Afrera water.

## 2.3 Kenya

There are many geothermal resources in Kenya, most of them in the rift zone where high subsurface temperatures exist due to the young volcanic activity. Kenya's geothermal capacity growth during the period 2015 to 2019 has been one of the fastest in the world, Huttrer(2020). Installations have totaled 218 MWe, coming from the Orpower4 (45 MWe) and the Olkaria V (173.2 MWe) stations. Current total installed capacity is 865 MWe, which comprises 29% of the national capacity. So far, more than 380 wells have been drilled in several parts of the rift zones. The location of the geothermal resources and young volcanoes is shown in Figure 3.

For this work, the geothermal waters were selected from four main studies carried out by Kamondo (1988) in several areas of the Kenyan Rift System (KRS), by Omenda (1998) in the Olkaria field, by Cioni *et al.* (1992) in the Lake Bogoria area, and by Sekento (2012) in the Menengai field.

The Olkaria high-temperature geothermal field, about 100 km NW of Nairobi, is located within the central Kenya segment of the East African Rift System (Fig. 3). It is the greatest geothermal field in Africa to generate electricity. The geothermal area is characterized by Quaternary volcanism of silicic composition of which the youngest is of Holocene age. The rock outcrops are dominated by comendite rhyolites and pyroclastics while in the subsurface are trachytes, basalts, rhyolites and tuffs. Geothermal manifestations include fumaroles, hot-springs and hot grounds. Numerous wells have been drilled to depths of 1000-2600 m and some have encountered temperatures of more than 300°C. The fluid chemistry at Olkaria broadly falls into two types: neutral pH-chloride and bicarbonate-rich waters. High enthalpy, neutral-chloride waters occur dominantly in the area east of Olkaria Hill, and lower enthalpy, bicarbonate-rich waters occur exclusively in the reservoir to the west of Olkaria Hill. For this study, we have selected geothermal waters from 4 wells (301, 305, 306 and 709) reported by Omenda (1998).

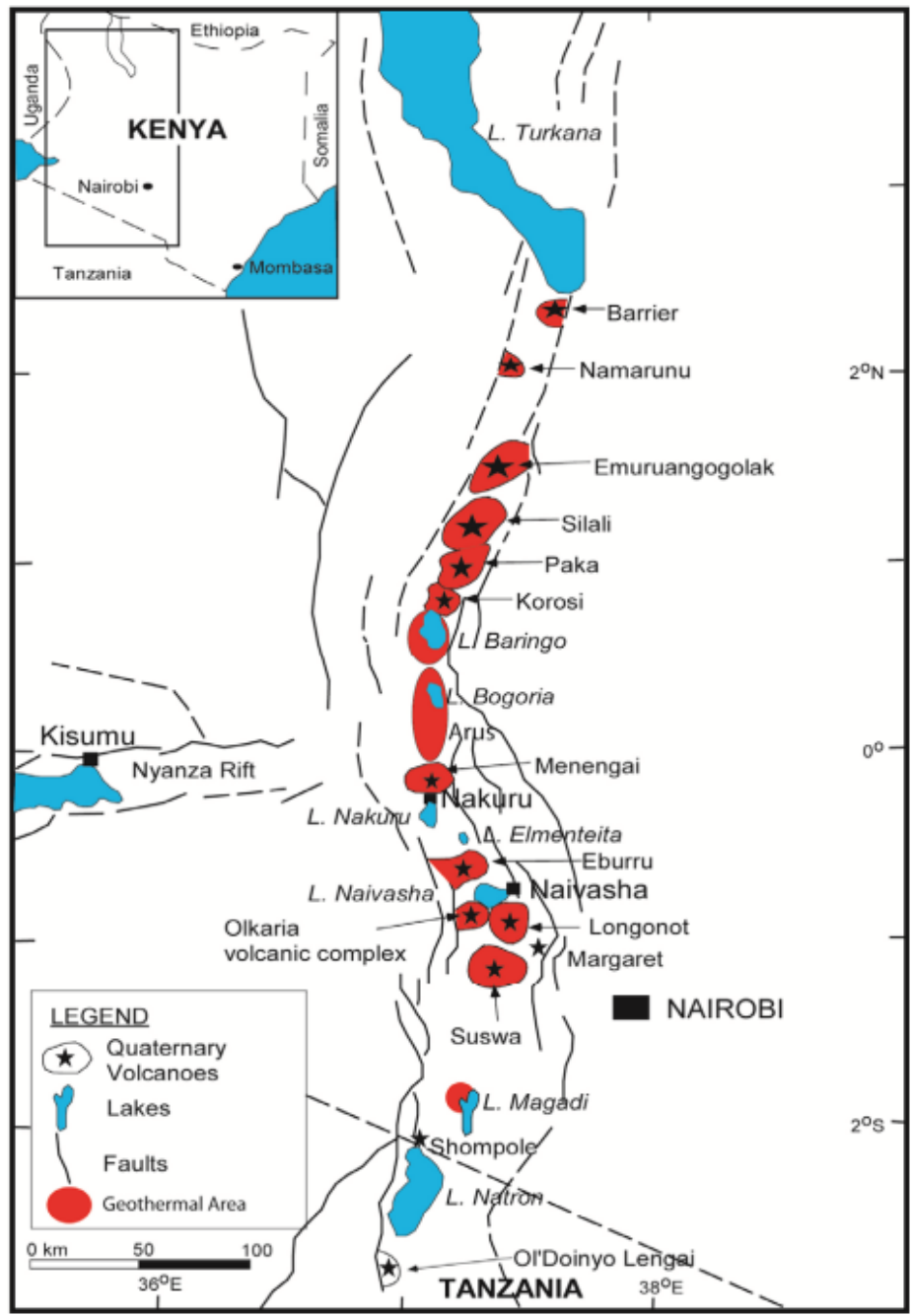

**Figure 3: Location map of the Kenyan Rift Valley showing the main geothermal areas, faults and quaternary volcanoes (extracted from Omenda and Teklemariam, 2010).**

The Lake Bogoria is found approximately 50-60 km north of the active Menengai volcano, within the main active branch of the KRS (Fig. 3). It is a closed-basin alkaline saline lake typical of African Rifts. A large number of boiling springs and fumaroles are located along the southern half of its shores. Menengai is a major Quaternary central volcano located in the KRS. It hosts one of the high-temperature geothermal fields located in the KRS (Fig. 3). The Menengai caldera is typified by complex tectonic activity associated with the rift triple junction. This is a zone at which the failed rift arm of the Nyanza rift joins the main Kenyan Rift (Fig. 3). The hydrothermal activity in Menengai geothermal field is manifested in the form of fumaroles, warm/ambient temperature boreholes, and hot to warm altered grounds.

The highest surface discharge temperature recorded around the Lake Bogoria is 98.9°C. Relevant hydrogeochemical data indicate that all the boiling springs have an extreme Na-HCO$_3$ composition, with high pH (from 8.0 to 9.8) and high salinity. Given that the thermal springs around the Lake Bogoria discharge mixtures of deep geothermal water, lake water (Na-HCO$_3$ type, TDS of 135000 mg/l, and pH close to 10), local groundwater, and steam, at different proportions. We selected the chemical compositions of the fluid samples 215-C, 324-C, 324-D, 325-D and 326-D successfully reconstructed by Cioni *et al.* (1992), using a mixing-boiling model. An equilibrium temperature close to 170°C was estimated for the deep geothermal reservoir using chemical geothermometers and modelling.

For the waters from the Menengai geothermal field, we selected the chemical compositions of fluid samples discharged from the well MW-01, given by Sekento (2012). These waters are Na-HCO$_3$ type and have pH values close to 9. The reservoir temperature was estimated at about 210°C using the quartz, Na-K and Na-K-Ca chemical geothermometers.

The geothermal waters from the Magadi and Silali areas were selected from the study carried out by Kamondo (1988). The Magadi hot springs occur around the margin of the Lake Magadi. The thermal waters are of alkaline Na-HCO$_3$ type and have a salinity in the range of 30-45 g/l. The hottest discharge recorded was 85°C. A reservoir temperature of about 100°C was estimated using the silica-chalcedony and Na-K geothermometers on the waters with TDS values lower than 40 g/l. The three waters with TDS values higher than 40 g/l were not selected, because their higher pH values (from 9.56 to 9.65), higher Na and Cl concentrations and lower silica and lithium concentrations suggest a mixing with the water of the Lake Magadi.

In the Silali area, only the Kapendo Na-HCO$_3$ thermal waters with a discharge temperature in the range of 27-51°C, TDS values up to 3.7 g/l and pH of 7.5-8.0, were selected for this study. The Na-K and silica geothermometers indicate a reservoir temperature of about 120°C, Kamondo(1988).

## 3. Results and discussion

### 3.1 Results

Following our literature review, we can note that the required geochemical data are the most abundant for the Republic of Djibouti. If the data relative to the Li concentrations are relatively numerous, very few data exist for the other trace elements. Only some data were found for F, Sr, Rb, and Cs. No sufficient representative data were obtained for Fe, Mn and W. Moreover, the number of geochemical data relative to waters from hot springs is much higher than for waters discharged from geothermal wells for which temperature values are measured at bottom-hole (Hanlé area in the Republic of Djibouti, Tendaho and Aluto-Langano areas in Ethiopia, Olkaria and Menengai areas in Kenya). The geochemical data obtained from the literature review are reported in Table 2.

For the geothermal areas represented by numerous hot waters with similar chemical compositions, such as those of the Sakalohl-Hallol, Hanlé, Lake Abhe in the Republic of Djibouti, Aluto-Langano in Ethiopia, Lake Bogoria, Silali and Magadi in Kenya, only the mean concentrations of elements and the respective standard deviations were considered in order to simplify and improve their presentation in the illustrations.

**Table 2: Chemical composition of the geothermal waters selected for this study from the literature review.**

| Geothermal area | Fluid sample | $T_{reservoir}$ °C | pH | TDS g/l | Na mg/l | K mg/l | Ca mg/l | Mg mg/l | Cl mg/l | HCO3 mg/l | SO4 mg/l | SiO2 mg/l | F mg/l | Li mg/l | Sr mg/l | Rb mg/l | Cs mg/l | Fe mg/l | Mn mg/l | Reference |
|---|---|---|---|---|---|---|---|---|---|---|---|---|---|---|---|---|---|---|---|---|
| **Republic of Djibouti** | | | | | | | | | | | | | | | | | | | | |
| Sakalol - Harralol | Mean Cluster C1 | 110 | 7.21-8.95 | 1.0 | 380 | 14.7 | 10.8 | 2.82 | 366 | 197 | 164 | 92.7 | 4.95 | 0.058 | 0.062 | | | | | Awaleh et al. (2017) |
| Sakalol - Harralol | Mean Cluster C2 | 110 | 7.07-8.65 | 3.5 | 1258 | 40.5 | 45.0 | 6.37 | 1766 | 120 | 273 | 95.8 | 2.62 | 0.096 | 0.280 | | | | | |
| Sakalol - Harralol | Mean Cluster C3 | 110 | 7.43-8.40 | 5.3 | 1844 | 50.1 | 112.0 | 13.7 | 2676 | 68 | 493 | 107 | 4.72 | 0.119 | 0.640 | | | | | |
| Sakalol - Harralol | Mean Cluster C4 | 140 | 7.15-7.16 | 13.3 | 3653 | 119 | 1020 | 12.9 | 7073 | 3.16 | 399 | 123 | 2.06 | 1.03 | 15.9 | | | | | |
| Hanle - Galafi | Mean Cluster | 145 | 8.01-8.75 | 2.4 | 823 | 42.5 | 19.7 | 2.69 | 912 | 229 | 328 | 91.7 | 3.14 | 0.043 | 0.112 | 0.051 | 0.076 | | | Awaleh et al. (2020) |
| Hanle - Minkille | Mean Cluster | 110 | 7.62-8.86 | 2.1 | 704 | 22.7 | 20.6 | 3.27 | 838 | 171 | 275 | 83.2 | 2.89 | 0.028 | 0.181 | 0.021 | 0.066 | | | Personal data (1988) |
| Ncinle | G1 | 140 | 7.85 | 0.75 | 136 | 5.5 | 15.4 | 2.28 | 138 | 272 | 122 | 57.7 | | 0.017 | 0.578 | 0.007 | 0.011 | | | Personal data (1988) |
| Ncinle | G2 | 140 | 7.88 | 1.2 | 340 | 18.4 | 15.8 | 2.14 | 383 | 114 | 201 | 129 | | 0.031 | 0.425 | 0.046 | 0.071 | | | |
| Lake Abhe | Mean Cluster SHC | 135 | 7.61-8.49 | 3.6 | 1099 | 31.3 | 228 | 1.35 | 1778 | 19.3 | 350 | 127 | 1.22 | 0.337 | 0.236 | | | 7.27 | 0.0068 | Awaleh et al. (2015) |
| Lake Abhe | Mean Cluster GHC | 135 | 8.11-8.79 | 2.1 | 591 | 17.0 | 157 | 1.34 | 908 | 19.74 | 60 | 43.7 | 1.78 | 0.244 | 0.226 | | | 7.97 | 0.0031 | |
| **Ethiopia** | | | | | | | | | | | | | | | | | | | | |
| Dofan-Fantale | Bulga 1 | 185 | 7.60 | 1.5 | 381 | 23.0 | 14.6 | 6.5 | 155 | 660 | 104 | 112 | 5.30 | 0.12 | | | | | | Pürshel et al. (2013) |
| Gergede-Sodere | Sodere 1 | 185 | 6.80 | 2.7 | 687 | 36.5 | 13.6 | 13.5 | 210 | 1387 | 138 | 146 | 9.20 | 0.22 | | | | | | Pürshel et al. (2013) |
| Gergede-Sodere | Sodere 2 | 185 | 7.10 | 2.0 | 490 | 23.5 | 17.4 | 16.3 | 101 | 1118 | 107 | 130 | 5.60 | 0.15 | | | | | | |
| Tendaho | TD-1 | 270 | 6.07 | 1.9 | 462 | 51.0 | 8.2 | 0.10 | 747 | | 38.4 | 640 | 1.30 | 0.65 | 0.410 | 0.270 | 0.190 | | | AQUATER (1996) |
| Tendaho | Mean TD-2 | 250 | 8.10-8.40 | 2.2 | 601 | 63.6 | 16.1 | 0.030 | 887 | 12.2 | 149 | 492 | 1.32 | 0.78 | 0.618 | 0.428 | 0.174 | 0.010 | 0.0020 | |
| Tendaho | Mean TD-4 | 250 | 8.12-8.42 | 2.3 | 663 | 86.9 | 7.9 | 0.035 | 900 | 19.9 | 124 | 488 | 0.99 | 1.08 | < 0.1 | 0.955 | 0.1333 | 0.078 | 0.00233 | |
| Aluto-Langano | Mean-springs | 185 | 7.60-9.10 | 2.6 | 764 | 41.0 | 2.3 | 0.67 | 361 | 1098 | 170 | 144 | 27.1 | 0.445 | | | | | | Pürshel et al. (2013) |
| Aluto-Langano | LA-4 (1987) | 260 | 8.40 | 3.9 | 1015 | 138 | 3.8 | 0.40 | 671 | 1647 | 131 | 339 | 27.8 | 1.30 | | | | | | |
| Other areas | Hertale | 160 | 7.35 | 0.97 | 238 | 16.0 | 8.0 | 5.0 | 118 | 403 | 78 | 101 | 1.80 | 0.030 | | | | | | Endeshaw (1988) |
| Other areas | Lake Abaya 6 | 260 | 10.00 | 4.6 | 1315 | 200 | 1.0 | 0.50 | 765 | 1821 | 102 | 428 | 43.3 | 1.38 | | | | | | Minissale et al. (2017) |
| Other areas | Lake Abaya 8 | 260 | 7.20 | 2.1 | 500 | 50.0 | 20.0 | 14.0 | 48 | 1286 | 12.0 | 146 | 13.0 | 0.580 | | | | | | |
| Other areas | Lake Afrera - 26 | 180 | 7.40 | 14.1 | 3208 | 250 | 1820 | 8.0 | 7952 | 37 | 176 | 92.0 | 1.00 | 0.800 | | | | | | |
| Other areas | Dallol | 110 | 7.50 | 47.6 | 13135 | 600 | 3720 | 680 | 28200 | 29 | 1167 | 99.0 | 1.00 | 0.300 | | | | | | |
| Other areas | Bilen | 140 | 6.90 | 1.00 | 200 | 18.0 | 25.0 | 10.0 | 64 | 537 | 37 | 107 | 3.30 | 0.020 | | | | | | |
| Other areas | Filweha | 100 | 7.85 | 1.7 | 450 | 85.0 | 2.0 | 2.0 | 187 | 824 | 76 | 86 | 7.00 | 0.010 | | | | | | |
| Other areas | Abaya 6 | 260 | 9.60 | 4.0 | 1290 | 240 | 0.20 | 0.10 | 711 | 1065 | 104 | 204 | 43.0 | 1.80 | | | | | | |
| Other areas | Bilate | 180 | 7.40 | 1.0 | 280 | 22.0 | 8.0 | 7.0 | 45 | 580 | 11.0 | 126 | 18.0 | 0.070 | | | | | | |
| Other areas | Dimtu well | 150 | 8.00 | 0.90 | 268 | 21.0 | 4.9 | 0.28 | 35 | 549 | 18.6 | 98.0 | 22.8 | 0.030 | | | | | | |
| **Kenya** | | | | | | | | | | | | | | | | | | | | |
| Olkaria | 709 | 300 | | 2.6 | 623 | 140 | 0 | 0 | 483 | 713 | 75 | 563 | 95 | 1.00 | | | | | | Omenda (1998) |
| Olkaria | 305 | 300 | | 2.2 | 453 | 87 | 0 | 0 | 466 | 520 | 44 | 597 | 33 | 1.40 | | | | | | |
| Olkaria | 301 | 275 | | 3.8 | 923 | 138 | 1.5 | 0.20 | 102 | 2200 | 93 | 333 | 66 | 2.40 | | | | | | |
| Olkaria | 306 | 250 | | 3.3 | 782 | 89 | 0.1 | 0 | 184 | 1724 | 60 | 431 | 37 | 0.80 | | | | | | |
| Lake Bogoria | Mean 215C, 324C, 324D, 325D, 326 D | 170 | 8.01-8.94 | 6.1 | 1774 | 86.7 | | | 623 | 3412 | 59.8 | 167 | 86.6 | 0.334 | | | | | | Cioni et al. (1992) |
| Silali | Mean 55, 56, 57, 58 | 120 | 8.10-8.30 | 3.3 | 1021 | 22.5 | 2.11 | 0.65 | 210 | 2025 | 81.5 | | | 0.055 | | | | | | Kamondo (1988) |
| Magadi | Mean 16, 17, 18, 19, 20, 21, 22 | 100 | 8.82-9.47 | 36.6 | 10549 | 165 | 1.0 | | 5450 | 20244 | 159 | 88 | | 0.923 | | | | | | Kamondo (1988) |
| Menengai | Mean MW-01 | 215 | 9.20 | 7.5 | 4314 | 275 | 2.03 | 0.18 | 675 | 7205 | 257 | 311 | 119 | 2.56 | | | 0.545 | | | Sekento (2012) |

For the reservoir temperatures estimated using chemical geothermometers applied on thermal waters, estimations given in the literature were sometimes modified. For example, in the Republic of Djibouti, where most of the geothermal areas indicate temperature estimations lower than 150°C, we have considered that the Na-K thermometric relationships empirically calibrated by Arnorsson et al. (1983) from Icelandic geothermal wells located in basaltic rocks in the range of 25-250°C or that proposed by Michard (1979), which gives similar results, were the most appropriate tools to estimate reservoir temperatures. The Na-K thermometric relationships determined by Fournier (1979) and Giggenbach (1988) are more appropriate for temperatures higher than 150°C.

For the silica geothermometer use, Arnorsson (1975) established that from temperatures measured in drillholes, the amount of dissolved silica in thermal waters at depth in the low-temperature hydrothermal areas in Iceland is governed by the solubility of chalcedony, when temperatures are below 110°C. At temperatures above about 180°C (160°C in Michard, 1979), the solubility of quartz governs the amount of dissolved silica in the water. Dissolved silica fits neither chalcedony nor quartz solubility distinctly in the temperature interval 110 to 180°C. However, Arnorsson et al. (1983) empirically calibrated the chalcedony thermometric relationship from Icelandic geothermal wells located in basaltic rocks in the range of 25-180°C and considered that equilibrium with chalcedony was approached at temperatures below about 180°C in Iceland. This thermometric relationship gives similar estimations than that defined by Fournier (1977). At this range of temperature, the estimations given using quartz and chalcedony thermometric relationships are different of about 25°C.

The K-Mg thermometric relationship defined by Giggenbach *et al.* (1983) can be also a useful tool for the low-temperature waters, whereas the Na-K-Ca thermometric relationship determined by Fournier and Truesdell (1973) often gives higher estimations of temperatures. Using these thermometric relationships, we propose a mean estimation of reservoir temperature of:

- 110°C for the geothermal waters of the clusters C1, C2 and C3 and 140°C for those of the cluster C4 from the Sakalol-Harralol area (instead of 143°C in the literature);
- 110°C and 145°C, respectively, for the geothermal waters of a first group constituted of the Minkillé, Dahotto, Daggirou, Oudgini, εagna springs and well H1, and a second group represented by the Galafi, Boukboukto, εasa Mayeb, Sâgallé, Dâli, εase-moydo and εaddara springs, from the Hanlé area (instead of 145°C for all the thermal springs in the literature).

For the two groups of geothermal waters from the Lake Abhe, we are in agreement with the mean reservoir temperature of 135°C given in the literature, Awaleh *et al.*( 2015).

### 3.2 Na-Li thermometric relationships

All the different existing Na-Li thermometric relationships are reported in Sanjuan *et al.* (2014). However, for this study, which is particularly focused on volcanic environment from a geological point of view, only two thermometric relationships were selected. The first one is that determined by Sanjuan *et al.* (2014), using dilute waters collected from wells located in different High-Temperature (200-325°C) volcanic geothermal areas of Iceland (Krafla, Námafjall, Nesjavellir and Hveragerdi) for which the reservoir temperature values were measured. This relationship can be expressed as follows:

$$T(K) = 2002/[\log\left(\frac{Na}{Li}\right) + 1.322]$$

where Na and Li are the aqueous concentrations of these elements given in mol/l.

The second empirical and statistical relationship is that defined by Fouillac and Michard (1981), based on numerous data of saline waters (Cl concentrations $\geq$ 0.3 mol/l) obtained from several geothermal fields mainly located in volcanic and crystalline areas:

$$T(K) = 1195/[\log\left(\frac{Na}{Li}\right) - 0.13]$$

where Na and Li are the aqueous concentrations of these elements given in mol/l.

The precision of the estimated temperatures for both thermometric relationships was statistically evaluated to ± 20°C (Sanjuan *et al.*, 2014).

The geochemical data obtained from the literature review (Table 2) are illustrated in Figure 4, which represents the log (Na/Li) of the geothermal waters (with the Na and Li concentrations expressed in mol/l) as a function of 1000/T, where T in Kelvin is the reservoir absolute temperature (measured for most of the wells or estimated using chemical geothermometers for the thermal springs or some wells).

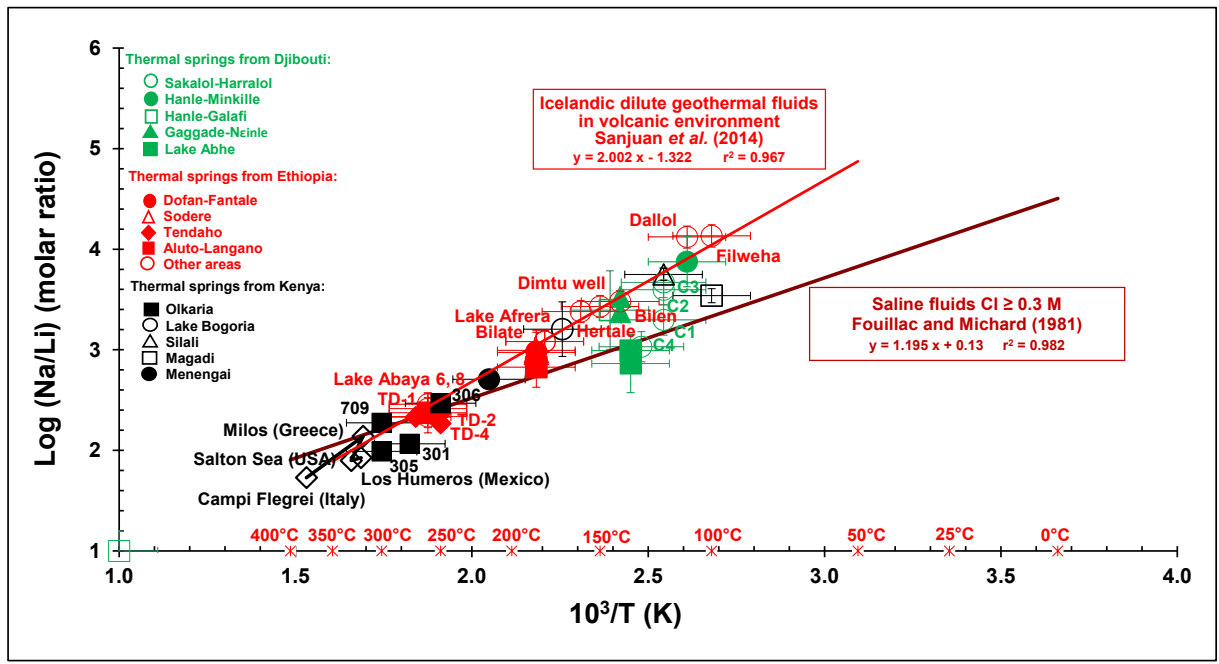

**Figure 4: Logarithm of Na/Li (molar ratio) *versus* 1000/T (reservoir temperature in K). The data from the literature review obtained in this study are compared to the Na-Li thermometric relationships defined by Sanjuan *et al.* (2014) for Icelandic HT dilute geothermal waters and by Fouillac and Michard (1981) for saline geothermal waters in volcanic and crystalline environments.**

Most of the geothermal waters selected for this study fit the Na-Li thermometric relationship defined by Sanjuan *et al.* (2014) relatively well. This suggests that the Na/Li ratios are probably controlled by an equilibrium reaction between, at least, K-feldspars, quartz, micas, albite, and illite minerals, Sanjuan *et al.*(2014). The Li concentrations of the geothermal waters could mainly come from the biotite dissolution during the alteration of the basaltic rocks, but other minerals could also contribute. A new thermodynamic approach using Li-minerals as that carried out by Boschetti (2022) could help to better determine the main Li-carrier minerals.

We can also see this thermometric relationship, which had been defined from Icelandic dilute geothermal waters, may be used for saline fluids, including brines from the Dallol area in Ethiopia. This had been already observed by Sanjuan *et al.* (2022) for saline waters from volcanic areas in Italy (Phlegrei Campi), Greece (Milos) and USA (Salton Sea), which are also reported in Figure 4. Consequently, the water salinity seems to be a parameter less influential than the nature of the reservoir rock or minerals on the use of the Na-Li thermometric relationship.

The geothermal waters of the clusters C1 and C4 from the Sakalol-Harralol area and from the Lake Abhe area in the Republic of Djibouti, and those from the Magadi area in Kenya rather fit the Na-Li thermometric relationship defined by Fouillac and Michard (1981) for saline waters (Fig. 4). This result is relatively surprising for the waters of the cluster C1 from the Sakalol-Harralol area and from the Lake Abhe area because these waters are not saline, but it is in agreement with the previous conclusion that water salinity is not the most influential parameter.

Some geothermal waters from the Hanlé plain, which were sampled and analyzed twice in the study of Awaleh *et al.* (2020), indicate very different Li concentrations whereas their chemical compositions are similar. It is difficult to explain these discrepancies, but for these waters, one of the two analyses at least fitted the Na-Li thermometric relationship determined by Sanjuan *et al.* (2014). Only these analyses were considered in the calculation of the mean concentrations of Li for the two groups of Hanlé geothermal waters.

In addition to the Icelandic and East-African geothermal fields, the Na-Li thermometric relationship was also successfully applied to dilute waters discharged from geothermal wells of the Los Humeros geothermal field in volcanic environment during the GeMex European project, Sanjuan *et al.*(2019), as shown in Figure 4. These results suggest this thermometric relationship can be still applied to other geothermal waters discharged in volcanic environment in the future and is a useful tool for geothermal exploration, which may be associated with the thermodynamic approach set up by Boschetti (2022) for more efficiency.

### 3.3 Other thermometric relationships (Na-Rb, Na-Cs, K-Sr, K-F)

Auxiliary chemical geothermometers such as Na-Rb, Na-Cs, K-Sr, K-Mn, K-Fe, K-F, K-W were determined by Michard (1990), from dilute waters discharged from granite reservoirs between 25 and 150°C in more than sixty areas from Europe (France, Italy, Spain, Bulgaria, Sweden). The corresponding thermometric relationships are reported in his paper.

Sanjuan *et al.* (2016a, b) also defined new Na-Rb, Na-Cs, K-Sr, K-Mn, K-Fe, K-F, K-W thermometric relationships using 20 hot natural brines from granite and sedimentary reservoirs, mainly located in the Upper Rhine Graben, France and Germany (70-230°C), apart two which were at Salton Sea, in the imperial Valley, USA (320-340°C).

In spite of the rare data relative to these trace elements in the literature review (Table 2), we could draw up four figures (Figs. 5A, 5B, 5C and 5D) representing the log ($K^2$/Sr ), log (Na/Rb), log (Na/Cs) and log (KxF) of the geothermal waters (with the Na, K, Sr, Rb, Cs and F concentrations expressed in mol/l), as a function of 1000/T, where T in Kelvin is the reservoir absolute temperature (measured for most of the wells or estimated using chemical geothermometers for the thermal springs or some wells).

In the Figure 5A, we can observe that the few geothermal waters from the Republic of Djibouti and Ethiopia, which are the only ones to have analyses of rubidium, fit the Na-Rb thermometric relationships determined by Sanjuan *et al.* (2016a, b) for brines relatively well. Once again, the water salinity seems to be a parameter of minor importance because most of the Djiboutian waters have relatively low TDS values.

In the Figure 5B, the alignment along the Na-Cs thermometric relationship determined by Sanjuan *et al.* (2016a, b) is less obvious, but possible. As for Li, the Rb and Cs concentrations could be associated with the dissolution of biotite, but additional bibliographical and experimental works must be carried out to be able to conclude.

In the Figure 5C, the geothermal waters from the Republic of Djibouti fit the K-Sr thermometric relationship defined by Sanjuan *et al.* (2016a, b) for brines more or less, whereas the geothermal waters from the Tendaho area are rather aligned along the K-Sr thermometric relationship defined by Michard (1990) for dilute geothermal waters in granite environment.

In the figure 5D, we can note that fluoride was the trace element for which we found the greater number of analyses. Overall, the geothermal waters with relatively low salinity fit the K-F thermometric relationship determined by Michard (1990) for dilute waters more or less and the geothermal brines from the Menengai, Lake Bogoria and Dallol areas are well-aligned along the K-F thermometric relationship defined by Sanjuan *et al.* (2016b). Only the geothermal waters with low salinity from the Filweha area also seems to fit this last relationship. The geothermal waters from the Tendaho area fit none of the thermometric relationships.

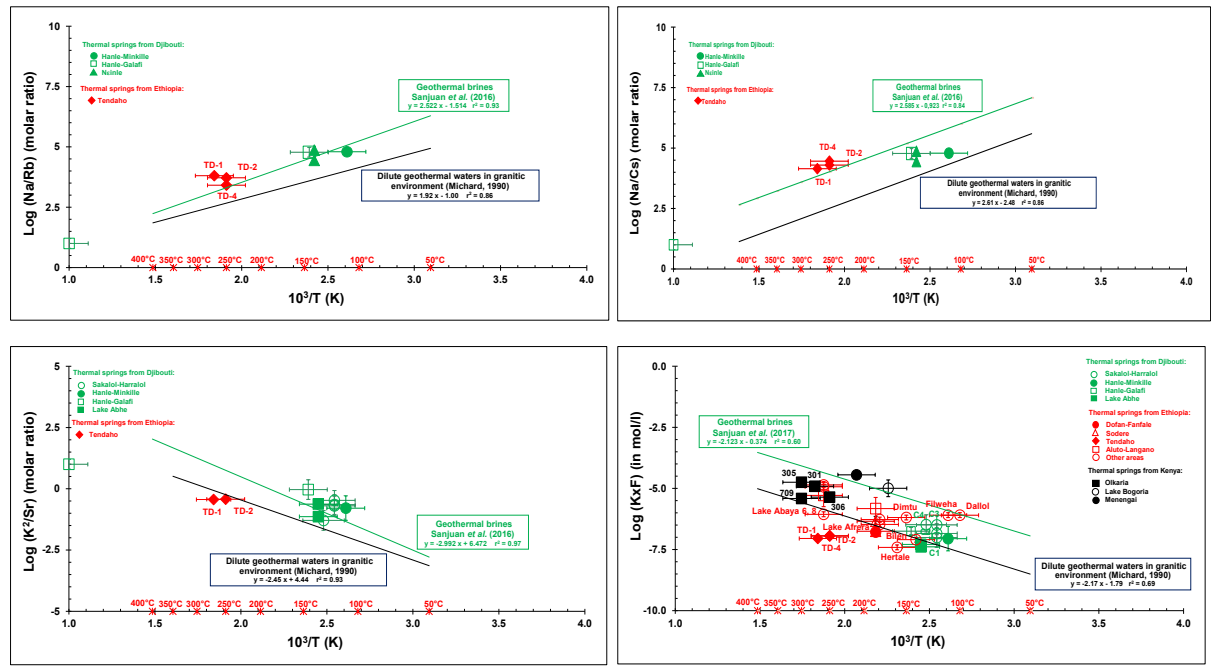

**Figure 5: A) Logarithm of Na/Rb (molar ratio) *versus* 1000/T (reservoir temperature in K). B) Logarithm of Na/Cs (molar ratio) *versus* 1000/T. C) Logarithm of K$^2$/Sr (molar ratio) *versus* 1000/T. D) Logarithm of KxF (in mol/l ) *versus* 1000/T. The data from the literature review obtained in this study are compared to the Na-Rb, Na-Cs, K-Sr and K-F thermometric relationships defined by Michard (1990) for dilute geothermal waters from granite environment and by Sanjuan *et al.* (2016a, b) for geothermal brines from granite and sedimentary reservoirs.**

To conclude, we note that more analyses of these trace elements would be necessary in the geothermal waters selected for our study to develop this type of auxiliary chemical geothermometers and understand the processes that control them, but these first results are rather promising.

## 4. Conclusions

This study has shown that the Na-Li thermometric relationship determined by Sanjuan *et al.* (2014) for dilute high-temperature geothermal waters from Iceland in volcanic environment can be an interesting tool for geothermal exploration in EARS areas. This thermometric relationship has given reliable estimations of reservoir temperatures for numerous low-, medium- and high-temperature geothermal waters from the Republic of Djibouti, Ethiopia and Kenya. Most of these waters have a relatively low salinity, but this parameter don't seem to be crucial on the control of the Na/Li ratio, contrary to the temperature and the nature of the reservoir rocks, because some brines can also fit this thermometric relationship.

For some dilute geothermal waters like those from one zone of the Sakalol-Haralol area and from the Lake Abhe area, in the Republic of Djibouti, the Na/Li ratio was rather controlled by the Na-Li thermometric relationship determined by Fouillac and Michard (1981). These relationships complement that defined by Sanjuan *et al.* (2014) for the geothermal fluids derived from seawater interacting with basalts at high-temperature ($\geq 160°C$) in Iceland (Reykjanes) and Djibouti (Asal-North Ghoubbet). For all these relationships, it is suggested that Li could be released by biotite dissolution. The new thermodynamic approach using Li-minerals as that carried out by Boschetti (2022) could help to better determine the main Li-carrier minerals. Concerning the use of the other auxiliary geothermometers, we have shown that very few geochemical data are available in the literature for geothermal waters from EARS areas. Only some interesting trends have been obtained for the Na-Rb, Na-Cs, K-Sr and K-F thermometric relationships, which need to be confirmed, but are rather promising. In the future, we encourage the community of geochemists to perform more analyses of trace elements such as F, Sr, Rb, Cs, Mn, Fe, and W in their studies, in order to develop additional tools for geothermal exploration.

*Acknowledgments:*

A major part of this work was carried out within the framework of the Workpackage WP9 - Geothermal African Atlas of the LEAP-RE project. This project has received funding from the European Union's Horizon 2020 Research and Innovation Program under Grant Agreement 963530.

# REFERENCES

AQUATER "Tendaho geothermal project" *Final report - Vol. 1*, (1996), 330 p.

Arnorsson, S. "Application of the silica geothermometer in low temperature areas in Iceland." *Am. J. Sci.*, 275 (1975), 763-774.

Arnorsson, S., Gunnlaugsson, E. and Svavarsson, H. "The chemistry of geothermal waters in Iceland. III. Chemical geothermometry in geothermal investigations." *Geochim. Cosmochim. Acta*, 47, (1983), 567-577.

Awaleh, M.O., Hoch, F.B., Boschetti, T., Soubaneh, Y.D., Egueh, N.M., Elmi, S.A., Jalludin, M., and Khaireh, M.A. "The geothermal resources of the Republic of Djibouti - II: geochemical study of the Lake Abhe geothermal field." *J. Geochem. Explor.*, 159, (2015), 129-147. https://doi.org/10.1016/j.gexplo.2015.08.011.

Awaleh, M.O., Boschetti, T., Soubaneh, Y.D., Baudron, P., Kawalieh, A.D., Dabar, O.A., Ahmed, M.M., Ahmed, S.I., Daoud, M.A., Egueh, N.M., and Jalludin, M. "Geochemical study of the Sakalol - Harralol geothermal field (Republic of Djibouti): evidences of a low enthalpy aquifer between Manda-Inakir and Asal rift settings." *J. Volcanol. Geotherm. Res.*, 331, (2017), 26-52. https://doi.org/10.1016/j.jvolgeores.2016.11.008.

Awaleh, M.O., Boschetti T., Adaneh A. E., Daouda, M.A., Ahmed M.M., Dabar O.A., Soubaneh, Y.D., Kawalieh, A.D., and Kadieh I.H. "Hydrochemistry and multi-isotope study of the waters from Hanlé-Gaggadé grabens (Republic of Djibouti, East African Rift System): A low-enthalpy geothermal resource from a transboundary aquifer." *Geothermics*, 86, (2020), 19 p. https://doi.org/10.1016/j.geothermics.2020.101805.

Boschetti, T. "A revision of lithium minerals thermodynamics: possible implications for fluids geochemistry and geothermometry." *Geothermics*, 98, (2022), 102286, 9 p. https://doi.org/10.1016/j.geothermics.2021.102286.

Cioni, R., Fanelli, G., Guidi, M., Kinyariro, J.K., and Marini L. "Lake Bogoria hot springs (Kenya): geochemical features and geothermal implications." *J. Volcanol. Geotherm. Res.*, 50, (1992), 231-246.

Endeshaw, A. "Current status of geothermal exploration in Ethiopia." *Geothermics*, 17, (1988), 477-488.

Fouillac, C., and Michard, G. "Sodium/Lithium ratios in water applied to geothermometry of geothermal reservoirs." *Geothermics*, 10, (1981), 55-70.

Fournier, R.O., and Truesdell, A.H. "An empirical Na-K-Ca geothermometer for natural waters." *Geochimica et Cosmochimica Acta*, 37, (1973), 1255-1275.

Foumier, R.O. "Chemical geothermometers and mixing models for geothermal systems." *Geothermics*, 5, (1977), 41-50.

Fournier, R.O. "A revised equation for the Na/K geothermometer." *Geotherm. Resour. Counc. Trans.*, 3, (1979), 221-224.

Giggenbach, W.F. "Geothermal solute equilibria, derivation of Na-K-Mg-Ca geoindicators." *Geochimica et Cosmochimica Acta*, 52, (1988), 2749-2765.

Giggenbach, W., Gonfiantini, R., Jangi, B.L., and Truesdell, A.H. "Isotopic and chemical composition of Parbati valley geothermal discharges, N.W. Himalaya, India." *Geothermics*, 12, (1983), 199-222.

Huttrer, G.W. "Geothermal power generation in the World 2015-2020 update report." *Proceedings World Geothermal Congress, Reykjavik, Iceland,* (2020), 17 p.

Kamondo, W.C. "Possible uses of geothermal fluids in Kenya." *Geothermics*, vol. 17, n° 2/3, (1988), 489-501.

Kharaka, Y.K., Lico, M.S., and Law, L.M. "Chemical geothermometers applied to formation waters, Gulf of Mexico and California Basins (abstract)." *A.A.P.G. Bull.*, 66, (1982), 588.

Kharaka, Y.K., and Mariner, R.H "Chemical geothermometers and their application to formation waters from sedimentary basins." *In: Naeser, N.D., McCulloch, T.H. (Eds.), Thermal History of Sedimentary Basins: Methods and Case Histories. Springer-Verlag*, New York, (1989), 99-117.

Michard, G. "Géothermomètres chimiques." *Bull. du BRGM (2ème série)*, Section III, n°2, (1979), 183-189.

Michard, G. "Behaviour of major elements and some trace elements (Li, Rb, Cs, Fe, Mn, W, F) in deep hot waters from granitic areas." *Chem. Geol.*, 89, (1990), 117-134.

Minissale, A, Corti, G., Tassi, F., Darrah,T.H., Vaselli, O., Montanari, D., Montegrossi, G., Yirgud, G., Selmo, E., and Tecluf, A. "Geothermal potential and origin of natural thermal fluids in the northern Lake Abaya area, Main Ethiopian Rift, East Africa." *J. Volcan. and Geoth. Research*, 336, (2017), 1-18. http://dx.doi.org/10.1016/j.jvolgeores.2017.01.012.

Omenda, P.A. "The geology and structural controls of the Olkaria geothermal system, Kenya." *Geothermics*, Vol. 27, n°1, (1998), 55-74.

Omenda, P.A., Teklemariam, M., "Overview of geothermal resource utilization in the East African Rift System." *Short Course V on Exploration for Geothermal Resources, UNU-GTP, GDC and KenGen, Kenya* (2010), 11 p.

Pürschel, M., Gloaguen, R., and Stadler, S. "Geothermal activities in the Main Ethiopian Rift: Hydrogeochemical characterization of geothermal waters and geothermometry applications (Dofan-Fantale, Gergede-Sodere, Aluto-Langano)." *Geothermics*, 47, (2013), 1-12. http://dx.doi.org/10.1016/j.geothermics.2013.01.001.

Sanjuan B., and Millot, R. "Bibliographical review about Na/Li geothermometer and Lithium isotopes applied to worldwide geothermal waters." *BRGM/RP-57346-FR report*, (2009), 58 p.

Sanjuan, B., Millot, R., Asmundsson, R., Brach, M., and Giroud, N. "Use of two new Na/Li geothermometric relationships for geothermal fluids in volcanic environments." *Chem. Geol.*, 389, (2014), 60-81.

Sanjuan, B., Millot, R., and Dezayes, Ch. "Three new auxiliary chemical geothermometers for hot brines from geothermal reservoirs." *Abstract: Goldschmidt Conference*, (2016a), Yokohama, Japan, 1 p.

Sanjuan, B., Gal, F., Millot, R., Dezayes, Ch., Jirakova, H., Frydrych, V., Nawratil de Bono, C., Martin F. "Chemical geothermometers and tracers." *Final IMAGE-D7.03 report*, (2016b), 74 p.

Sanjuan, B, Gal, F., and Cuevas Villanueva, R.A. "Developments of auxiliary chemical geothermometers applied to Los Humeros and Acoculco high-temperature geothermal fields (Mexico)" *In GeMex Deliverable D4.3 on geochemical characterization and origin of cold and thermal fluids:* Chapter 3, (2019), 55-100.

Sanjuan, B., Gourcerol, B., Millot, R., Rettenmaier, D., Jeandel, E., and Rombaut, A. "Lithium-rich geothermal brines in Europe: an up-date about geochemical characteristics and implications for potential Li resources". *Geothermics*, 101, (2022), 18 p., 102385. https://doi.org/10.1016/j.geothermics.2022.102385.

Sekento, L.R. "Geochemical and isotopic study of the Menengai geothermal field, Kenya". *Final report*, n°31, (2012), Geothermal training programme, UN University, 24 p.

Zan, L, Gianelli, G., Passerini, P., Troisi, C., Hagas A.O. "Geothermal exploration in the Republic of Djibouti: thermal and geological data of the Hanlé and Assal areas." *Geothermics*, vol. 19, n°6, (1990), 561-582.